\documentstyle[sprocl]{article}

\input{psfig}
\def\be{\begin{equation}}
\def\ee{\end{equation}}
\def\bea{\begin{eqnarray}}
\def\eea{\end{eqnarray}}
\begin{document}

\title{Nucleus-Nucleus Collisions at Highest Energies}

\author{ M.~Bleicher, N.~Amelin, S.~A.~Bass, M.~Brandstetter, A.~Dumitru,
C.~Ernst, L.~Gerland,
J.~Konopka, C.~Spieles, H.~Weber, L.~A.~Winckelmann,   
H.~St\"ocker and W.~Greiner}

\address{Institut f\"ur Theoretische Physik\\
Johann Wolfgang Goethe Universit\"at\\
    60054 Frankfurt am Main, Germany}


\maketitle\abstracts{The microscopic phasespace approach {\small URQMD} is 
used to investigate the stopping power and particle production in heavy 
systems at SPS and RHIC
energies. We find no gap in the baryon rapidity distribution even at RHIC.
For CERN energies {\small URQMD} shows a pile up of baryons and a supression of
multi-nucleon clusters at midrapidity. 
}

nucl-th/9605044

\section{Motivation}

One of the main aims of relativistic heavy ion collisions at collider
energies is to discover if the individual hadrons dissolve into a gas
of free quarks and gluons (quark-gluon-plasma, QGP) in the extremely
compressed and heated hadronic matter. This may happen
in line with a transition into the chiral symmetric phase which
modificates most hadron masses drastically. The achievable energy 
deposition depends on the amount of stopping of the colliding nuclei. 

\section{The {\small URQMD} Model}

The Ultrarelativistic Quantum Molecular 
Dynamics ({\small URQMD})\cite{URQMD},
is used to analyze the physics of
the excitation function of hadronic abundances, stopping and flow.
This framework bridges with one model consistently the entire available 
range of
energies from below SIS to CERN, even for the heaviest system Pb+Pb.  
{\small URQMD} is a hadronic transport model including strings.
Its collision term contains 50 different baryon species
(including nucleon, delta and hyperon resonances with masses up to 2 GeV)
and 25 different meson species (including strange meson resonances), which
are supplemented by their corresponding antiparticle
and all isospin-projected states.

\section{He-He Collisions at ISR}

\begin{figure}
\centerline{\psfig{figure=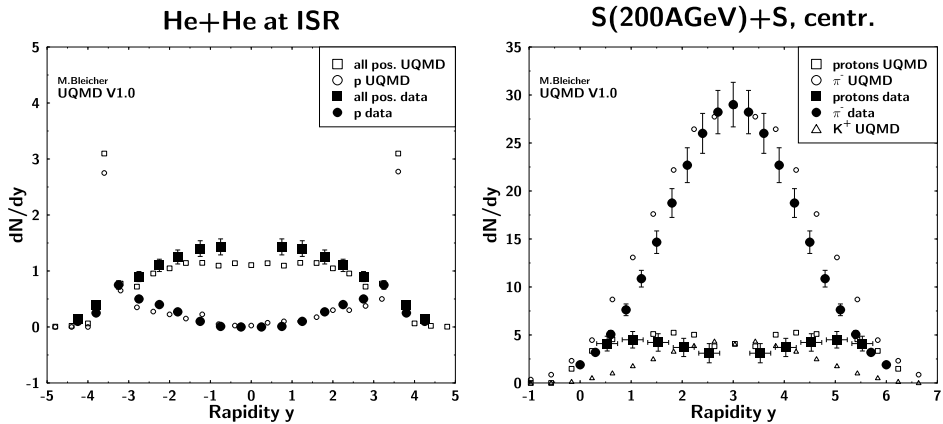,width=13cm}}
\caption{\label{fig:he31he}Left:Rapidity distribution of protons and
positively charged particles for the reaction He+He
at $\sqrt{s}=31$~AGeV compared to the data \protect\cite{otterlund}.
Right: Rapidity distribution of protons and
negatively charged pions for central S+S reactions
at $E_{lab}=200$~AGeV compared to the data \protect\cite{stroebele}.}
\end{figure}

In general it appears to be an intricate problem 
to describe stopping behaviour of baryons and pion production
within one theoretical frame at very high energies. If one wants to do 
LHC calculations it may be necessary to include
multi-string-excitation as it is done in dual parton approach. 
To demonstrate the ability of {\small URQMD} to model a nucleus-nucleus
collision even at the today highest available bombarding energies for
heavy particles, we compare the calculated He+He collision at ISR with
data \cite{otterlund} as shown in Fig.\ref{fig:he31he} (left).

It is not surprising, that such a light system as helium is totally
transparent. A baryon free area of 3 units in rapidity is produced.
{\small URQMD} and the data agree well, the calculated produced particle yield may
increase by 15\% if one also simulates gluon jet events \cite{tom}, which is not
included yet. 

\section{SPS Energy Regime}

The dominant reaction mechanism
in the early stage of a reaction
 is the excitation
of collision partners to
resonances or strings\cite{URQMD}.
Then secondary interactions, i.~e.
the annihilation of produced mesons on
baryons, lead also to the formation
of $s$ channel resonances or strings, 
which may explain the strangeness enrichment \cite{RQMD2} and
(for masses larger than 3$m_N$) allow for
$\overline{N}N$ creation\cite{aja94}.
The escape probability for $\bar p$'s
from the exploding nuclear matter enters via the
free $N\overline{N}$ annihilation cross
section. For central events of Pb+Pb at SPS approx. 85\% of the produced
anti-baryons are annihilated during the reaction.
\begin{figure}
\centerline{\psfig{figure=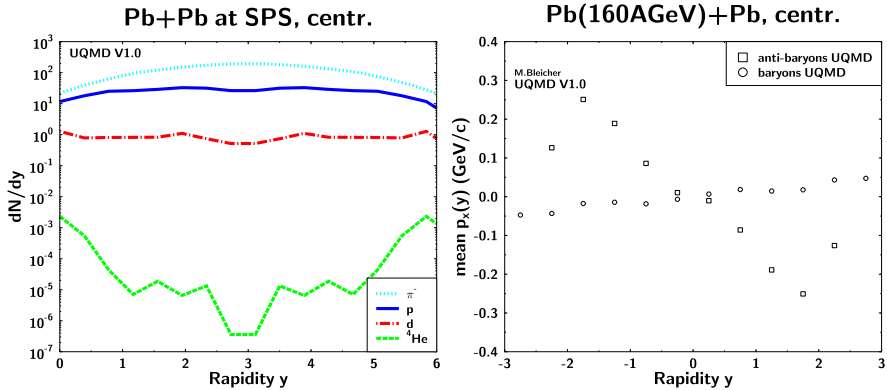,width=12cm}}
\caption{\label{fig:sps}Left: Rapidity distribution of $\pi^-$, protons,
deuterons and ${}^4$He in central Pb+Pb reactions
at $E_{lab}=160$~AGeV. Right: Directed flow of baryons and
antibaryons for the same system.}
\end{figure}

These two counter--acting effects may be
measured by the directed ``antiflow'' of antimatter.
The observable asymmetry
 for bouncing antimatter can be quantified by the 
mean $p_x$ vs. rapidity (Fig.\ref{fig:sps}, right).
The anti--flow of antibaryons appears to be
strongest for semicentral collisions, while for baryons
the maximum $p_x$ is at considerably smaller 
$b$--values. The latter is due 
to the pressure (i.~e. the EOS) the former one due to absorption and
geometry. 

Comparisons of {\small URQMD} calculations to data from SIS to SPS is documented
elsewhere\cite{bass}. Good agreement of baryon and meson production and dynamics
has been achieved. An impression is given in Fig.\ref{fig:he31he} (both) - 
protons as well
as pions and kaons are shown. 
\begin{figure}
\centerline{\psfig{figure=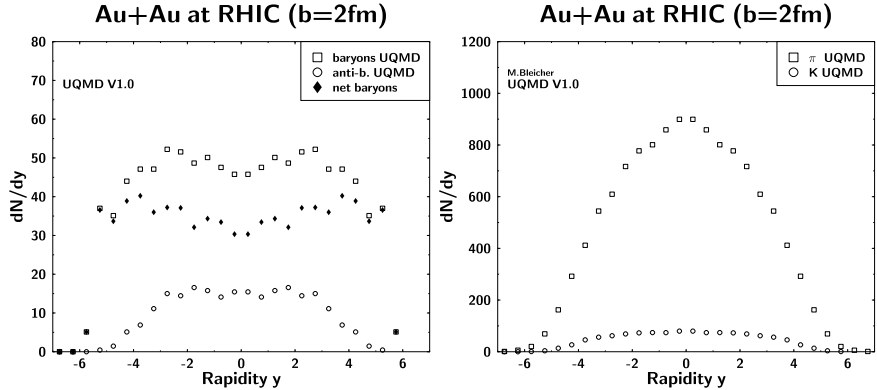,width=12cm}}
\caption{\label{fig:rhic}The system Au+Au at an energy of
$\sqrt{s}=200$AGeV (RHIC), central collisions selected.
Left:Rapidity distribution of baryons,
antibaryons and net-baryons. Right: Rapidity distribution of pions and 
kaons}
\end{figure}

Further insight into the collision geometry may be gained by looking at
composite particle probes. 
Since {\small URQMD} does not include the production of light
nuclei dynamically, cluster formation is added after strong freeze-out.
(Freeze-out means after the last strong interaction of the particle.)
We calculate the deuteron (helium) formation probability by 
projecting the
nucleon pair phasespace on the deuteron wave function via
the Wigner-function method as described in \cite{raffi}. 
Especially for high bombarding energies or exotic clusters it is
necessary to use this sophisticated method, since the complex phasespace
distribution of the cluster ingredients has to be taken into account
\cite{blubb}. The Hulth\'en
parametrization is used to describe the relative part of the
deuteron wavefunction.
The yield of deuterons is given by
\[ {\rm d}N_{d}={1\over 2}{3\over
4}\Big<\sum\limits_{i,j}\rho^{^{\rm W}}_{{d}}
(\Delta \vec{R},\Delta \vec{P})\Big>
{\rm d}^3(p_{i_{{p}}}+p_{j_{{n}}})\,. \]
The Wigner-transformed Hulth\'en wavefunction of the
deuteron is denoted by $\rho^{^{\rm W}}_{{d}}$.
The sum goes over all ${n}$ and ${p}$
pairs, whose relative distance ($\Delta \vec{R}$) and relative momentum
($\Delta \vec{P}$) are calculated in their rest frame at the earliest
time after both nucleons have ceased to interact. The
factors ${1\over 2}$ and ${3\over 4}$ account for the statistical spin and
isospin projection on the deuteron state. The calculation of the
high mass clusters is straight forward, e.g. by exchanging the Hulth\'en
parametrization of the d wavefunction with a 4-body harmonic oscillator
wavefunction\cite{raffi} to describe the ${}^4$He 
(See Fig. \ref{fig:sps} (left), 
$\pi^-$, protons, as well as deuterons and ${}^4$He are depicted).
In contrast to the pile up of protons at midrapidity, cluster production
is strongly supressed due to the high temperatures in the center of the
collisions. 

Calculations of $H^0$ ($\Lambda\Lambda$-clusters) for AGS and SPS
energies have also be performed in this framework \cite{spieles}.

\section{RHIC Estimates}

As shown above {\small URQMD} seems to be well suited for an estimate of stopping
power in Au+Au collisions at RHIC. However, we are well aware of the
fact that for certain observables (e.g. high $p_t$ components of
particle spectra) the current framework is not sufficient, since it does
not incorporate hard partonic scattering explicitely. 
Figure \ref{fig:rhic} (left) shows
the results of our calculation for gold-gold at $\sqrt{s}=200$~AGeV
(RHIC). The nuclei suffer a mean rapidity shift of more than 2 units of
rapidity. The mid-rapidity region is apparently not baryon free, in
contrast to some earlier expectations. Our results are similiar to
{\small RQMD} calculations\cite{tom}. 

We finally put our interest on the produced mesons. In
Fig.\ref{fig:rhic} (right) we show the rapidity distribution of kaons and
pions. As mentioned above at this high energies modifications of the
string fragmentation may be necessary. The inclusion of gluon jets as a
first step will increase the meson multiplicities by about 15\% \cite{tom}. 

\section{Conclusion}

At first some remarks on the validity of our calculation: It is believed
to be a fact that a quark-gluon-plasma is created at RHIC. While our
calculation does not assume a QGP, it certainly goes beyond the purely
hadronic scenario - strings are excited and quarks and diquarks are
subject to further interaction. From our calculation we  infer that the
interaction of leading quarks and diquarks dictate the stopping
behaviour of heavy ion collisions even at SPS energies.

\section*{Acknowledgments}
This work is supported by GSI, DFG and BMBF.
M.B. wants to thank R. Mattiello for many inspiring 
ideas and fruitful discussions.

\end{document}